# GitHub Marketplace for Automation and Innovation in Software Production

SK. Golam Saroar, Waseefa Ahmed, Elmira Onagh, Maleknaz Nayebi

*EXINES lab, York University*

**Abstract**

**Context:** GitHub, renowned for facilitating collaborative code version control and software production in software teams, expanded its services in 2017 by introducing GitHub Marketplace. This online platform hosts automation tools to assist developers with the production of their GitHub-hosted projects, and it has become a valuable source of information on the tools used in the Open Source Software (OSS) community.

**Objective:** In this exploratory study, we introduce GitHub Marketplace as a software marketplace by exploring the Characteristics, Features, and Policies of the platform comprehensively, identifying common themes in production automation. Further, we explore popular tools among practitioners and researchers and highlight disparities in the approach to these tools between industry and academia.

**Method:** We adopted the conceptual framework of software app stores from previous studies and used that to examine 8,318 automated production tools (440 Apps and 7,878 Actions) across 32 categories on GitHub Marketplace. We explored and described the policies of this marketplace as a unique platform where developers share production tools for the use of other developers. Furthermore, we conducted a systematic mapping of 515 research papers published from 2000 to 2021 and compared open-source academic production tools with those available in the marketplace.

**Results:** We found that although some of the automation topics in literature are widely used in practice, they have yet to align with the state of practice for automated production. We discovered that practitioners often use automation tools for tasks like "Continuous Integration" and "Utilities," while researchers tend to focus more on "Code Quality" and "Testing".

**Conclusion:** Our study illuminates the landscape of open-source tools for automation production. We also explored the disparities between industry trends and researchers' priorities. Recognizing these distinctions can empower researchers to build on existing work and guide practitioners in selecting tools that meet their specific needs. Bridging this gap between industry and academia helps with further innovation in the field and ensures that research remains pertinent to the evolving challenges in software production.

*Email address:* saroar@yorku.ca, waseefa@yorku.ca, eonagh@yorku.ca, mnayebi@yorku.ca (SK. Golam Saroar, Waseefa Ahmed, Elmira Onagh, Maleknaz Nayebi)

## 1. Introduction

Multi-sided markets are economic platforms that connect various distinct user groups through interdependent relationship networks where the success of one side positively impacts the others [16]. Software platforms such as app stores are quintessential examples of such markets. The growing visibility of these platforms is quite note-worthy. These platforms serve as digital infrastructure for digital ecosystems that foster open innovation and facilitate the distribution of software tools and applications while playing a pivotal role in the economic aspect of software distribution, underscoring the growing importance of multi-sided markets within the software distribution landscape. Notably, app store architectures, along with underlying application programming interface (API) architectures, play a crucial role in shaping the dynamics of these platforms. From the information systems perspective, the app stores can be seen as pivotal nodes in digital ecosystems, functioning as transaction platforms, hybrids, or boundary resources that align closely with innovation platforms like operating systems. This viewpoint clarifies their integral role in facilitating software distribution and fostering open innovation within the software market landscape [17, 35, 10]. In software Engineering, the success of the app store ecosystem hinges on the symbiotic relationship between the two groups: developers and end users [64, 83].

On the one hand, a thriving developer community uses app stores as centralized hubs to showcase and distribute their software applications, contributing to a richer app selection and benefiting from a broader reach without extensive marketing efforts. Conversely, end-users rely on app stores as convenient and secure marketplaces to discover and access applications tailored to their needs, enticing developers seeking a larger audience [35, 64]. GitHub Marketplace is a unique marketplace where, unlike mobile app stores, the users and the providers are both software developers (see Figure 1). GitHub Marketplace allows users to discover free or paid tools for automating the workflows [20] of their GitHub-hosted projects. This platform offers two variations of automation tools: *Apps* and *Actions*. While both facilitate the automation of various aspects of a given project, they each have their unique characteristics [24]. **GitHub Actions** are free tools designed for the automation of customized individual tasks within a workflow [21], including automation for CI/CD pipelines, building, testing, or deploying software projects. Alternatively, **GitHub Apps** can be free or paid with multiple paywalls depending on their functionality. Apps can be installed on user or organization accounts hosted on GitHub with explicit access to specified repositories [22]. Developers can classify their tools into a maximum of two distinct categories. These categories are pre-determined by GitHub Marketplace and describe a high-level objective of the tools.

Figure 1 demonstrates the elements of this GitHub Marketplace on an adopted conceptual model offered by Jansen and Bloemendal [35]. The actors in this model are GitHub Marketplace as the *owner of the platform*, the developers seeking automation tools as *end-users*, and the developers sharing and distributing automation tools as *providers*. The features that actors can interact with are shown in Figure 2 and are divided into *Characteristics* (not under the direct influence of GitHub Marketplace) and *Features & Policies* (directly influenced by GitHub Marketplace). Despite the long-standing interest of the software engineering community in automating development workflows and providing decision support, the GitHub Marketplace has not been subject to investigations until now. The automation tools in GitHub Marketplace (Apps and Actions) possess both technical (Code repositories) and non-technical attributes (user attributes such as star rating), indicating this platform would be a valid choice for analysis as suggested by Martin et al. [47]. We combined this perspective with the one by Harman

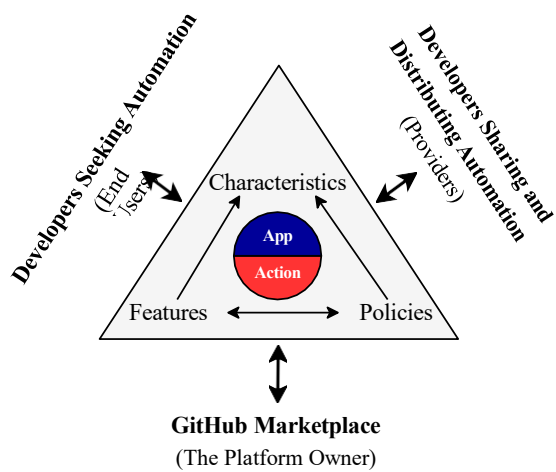

Figure 1: The conceptual model of GitHub Marketplace mapped to Jansen and Bloemendal [35] app store model. Developers use GitHub Marketplace to share automated production tools in the form of "Apps" and "Actions" with the other developers hosting their projects on GitHub.

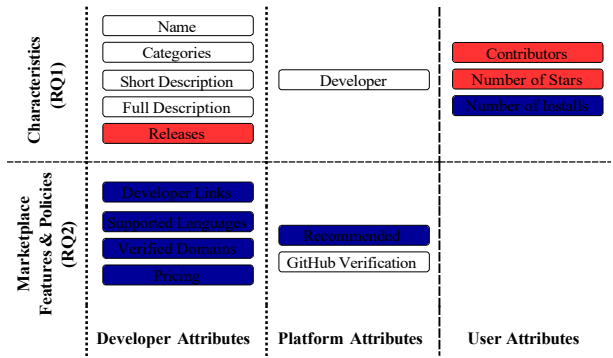

Figure 2: Characteristics, Features, and Policies surrounding the automated production tools on GitHub Marketplace [35, 47]. Attributes for Apps and Actions are color-coded as blue and red, respectively, and the attributes used by both are depicted as white.

et al. [30, 47] around different actors in Figure 2.

While extensive analyses have focused on mobile app stores to understand the dynamics between users and developers, the GitHub Marketplace remains relatively underexplored. Particularly noteworthy is the developer-to-developer exchange of automation tools. Exploring the GitHub Marketplace offers a unique opportunity to assess the alignment between cutting-edge and practical open-source software production tools. This exploration promises insights into evolving practices and innovations within the ecosystem of software development tools. GitHub Actions were first introduced as an alternative to CI/CD services for GitHub repositories [11]. Later, the Actions were extended to cover more than 30 different types of automation in the production pipeline. Leveraging this marketplace as a valuable data source, we aim to evaluate the alignment or divergence between research efforts (published studies) and practical applications (automation tools on GitHub Marketplace). This exploration contributes to understanding how the broad field of automated software production compares in academia and practice. In this context, we answer three research questions (RQs):

**RQ1:** What are the characteristics of GitHub Marketplace, and what automated production tools are distributed and used on the platform?
**Why and How:** As a first study on GitHub Marketplace, we introduce its characteristics through a comprehensive mining study. We collected a detailed dataset and conducted descriptive analysis, as well as statistical analysis when applicable. This also allowed us to understand the automated production tools on the platform both quantitatively and qualitatively.

**RQ2:** What features and policies govern the GitHub Marketplace, and what is their current status?
**Why and How:** To understand and introduce the features and policies of the marketplace in the first study of its kind, we adopted a mining approach, building upon **RQ1**. We cross-evaluated our observations with the existing documentation on policies and features whenever possible and provided and described the status quo qualitatively and quantitatively.

**RQ3:** How does the academic literature relate to practical tools distributed on the GitHub Marketplace for automated software production?
**Why and How:** To understand the gap between state-of-the-art and state-of-the-practice and evaluate the potential for mobilization, we performed a systematic literature analysis and mapped the papers retrieved from our systematic search to the categories and topics of automated production tools on GitHub Marketplace and compared their popularity.

In the following sections, we first discuss related works done in the field (section 2), followed by our methodologies in section 3, followed by a summary of our results in section 4 and discussion and implications of our study in section 5. We conclude our paper with a discussion on threats to the validity of our project in section 6 and a summary of our findings and conclusion in section 7.

## 2. Related Work

GitHub Marketplace as a distribution platform of tools for automated production nor as a software repository has not been discussed in the literature. However, in recent years, marketplaces have drawn much attention among researchers within the software engineering community.

Software marketplaces have been analyzed from the ecosystem perspective [35]. Messerschmitt [49] analyzed the influence of marketplace issues, such as related issues, on software project planning, considering the compromise between revenue and distribution cost on multiple platforms. Derave et al. [13] studied the business model of digital marketplaces by extending Digital Platform Ontology to understand the various business model choices made in digital marketplaces. Additionally, the study proposed a conceptual model for developing software platforms to assist developers with analyzing the influence of the business model decisions on specific platform functionality. Jazayeri et al. [36] identified the common features and interrelations of IT service markets to provide a comprehensive understanding of app markets. They performed a systematic literature review and created a reference model for IT service markets to be used for the creation and maintenance of the markets.

Mehrotra et al. [48] suggested a framework for improving marketplace characteristics. Hyrynsalmi et al. [33] studied multi-homing (hosting a tool on multiple platforms) in mobile software ecosystems to understand the impact of this strategy on mobile platforms. They found that mobile application ecosystems are generally single-homing markets. The developers of superstar applications commonly use a multi-homing strategy [34].

> *GitHub Marketplace is a platform where developers provide automation tools (Apps and Actions). This marketplace has never been researched in software engineering. This is despite the seven mining studies that were conducted on a random or a convenient sample of repositories that are using Actions to automate their processes.*

Marketplaces have also been evaluated through the mining lens. Krüger et al. [41] described a preliminary analysis of the Eclipse Marketplace with a view of looking at open marketplaces. They proposed to mine marketplace data to address questions such as who contributes to successful plug-ins and to identify leading developers and communities, leading to collaborations and new research directions. Li et al. [43] performed a performance and usability evaluation of five open-source IDE plugins that identify and report security vulnerabilities. Analysis of mobile app stores has been widely adopted since the study of Harman et al. [30], where they provided a framework for analyzing the app marketplace. Afterward, Chen et al. [7] proposed ARMiner, a novel app review mining framework to mine requirements for better market competition. They characterized the mobile app stores in the BlackBerry app store based on technical, customer, and business attributes. The systematic review by Martin et al. [47] summarized an extensive body of literature on mobile app stores. However, the look into the mobile app stores through the lens of the software business and as a two-sided market has been rather sparse [64, 62].

## 3. Research Methodology

The methodology used in this study consists of two main parts. In the first part, using empirical protocols, we retrieved and mined different metadata (attributes) for both GitHub Marketplace and the automation tools hosted on this platform to answer the first two research questions. We used statistical analysis to establish GitHub Marketplace as a software marketplace by identifying the platform's Characteristics, Features, and Policies (Figure 2). We followed the state-of-the-art methods [40, 39, 79, 67, 82, 77, 32, 3, 54, 53, 12]. A variety of analytical models for software and project management exists in software engineering research [70, 75, 45, 80, 76, 63, 31] which we build our study upon such studies [81, 66, 54, 60, 68, 55, 67, 56]. To answer the third research question, we performed a comprehensive literature analysis to extract the spectrum of software engineering research topics regarding open-source automation tools. Finally, we used the classification of the automation tools based on GitHub Marketplace's pre-defined categories as our main method to establish links between automation tools hosted on GitHub Marketplace and academically sourced ones [71, 59, 46, 58, 29, 73, 74, 61, 65, 69, 37, 72].

### 3.1. GitHub Marketplace Automated Production Tools (RQ1 & RQ2)

In this section, to answer the **RQ1** and **RQ2**, we extracted various attributes that define the GitHub Marketplace platform and the tools it hosts. To gather these attributes from the GitHub Marketplace, we used Selenium [85] (for automated loading of the JavaScript components) along with Scrapy [84] (as the main web page crawler). We noted that the marketplace is limited in displaying (hence automatically accessing) tools when the number of tools in a given category exceeds 1,000. While this was not problematic for App data (at the time of this study, there were less than 1,000 Apps in the marketplace), this restriction hindered us in extracting all Action data for a given category of GitHub Marketplace. To avert this restriction, we took advantage of the GitHub Marketplace's sorting algorithms view ('Best Match', 'Recently added', etc.). Each sorting algorithm displayed a different data set, allowing us to access additional Actions.

To scrape the data from the marketplace, we designed a set of knitted crawlers that started from the marketplace's main page and took each category as a new scrapping root. Having a set of scrapping roots, we visited each tool's web page on GitHub Marketplace to gather descriptive data about the tool. We observed that the Apps and Actions detail pages differ from each other, and developers provide different levels of detail about each tool. We parsed through the different HTML header layers for each tool's web page and uniformly gathered nine Action attributes and 12 App attributes across the marketplace, six of which are common between both Actions and Apps. We classified these attributes into three collections: *Developer Attributes* (defined by the developers when publishing their tools on the marketplace), *Platform Attributes* (defined by the marketplace), and *User Attributes* (defined by the end-user). Additionally, we divided these attributes into *Characteristics* (ones that GitHub Marketplace has no control over) and *Features & Policies* (ones that are under the control of the marketplace to some extent). Figure 2 includes details of these attributes divided into their corresponding classifications. The Action and App attributes are distinguished by the use of color (red for Actions, blue for Apps, and white for the common attributes) and discussed in detail in Sections 4.1 and 4.2.

## 3.2. Comparing State of the Art and Practice on Automated Production in Open Source (RQ3)

Once we understood the Characteristics, Features, and Policies of the GitHub Marketplace, our goal was to ascertain the comparability between open-source literature on software automation production and the tools utilized by practitioners in the marketplace, exploring this aspect in **RQ3**. To this end, we systematically gathered and mapped literature in such a way as to allow comparison with the GitHub tools.

### 3.2.1. Literature on Automated Production in Open Source

To answer **RQ3**, we conducted a systematic mapping study as suggested by Peterson et al. [78] to compare current trends in the open-source research community with the ones in practice. Systematic mapping, by its nature, is designed to conduct a comprehensive analysis of topics and trends in a specific field, making this an ideal approach for comparing the state-of-the-art with the state-of-the-practice. In our study and along with our three research questions, we used multiple mapping schema, including the categories available in GitHub Marketplace. Our goal was to directly compare state-of-the-art developments with current practices. While the GitHub tools are already categorized, an intuitive and natural approach for this study was to map the papers for a side-by-side comparison. We summarized an overview of our approach to systematic mapping, which resulted in the selection of 515 papers for this study in Figure 3.

**Source Selection:** Following previous studies, we gathered papers from five publication sources, namely

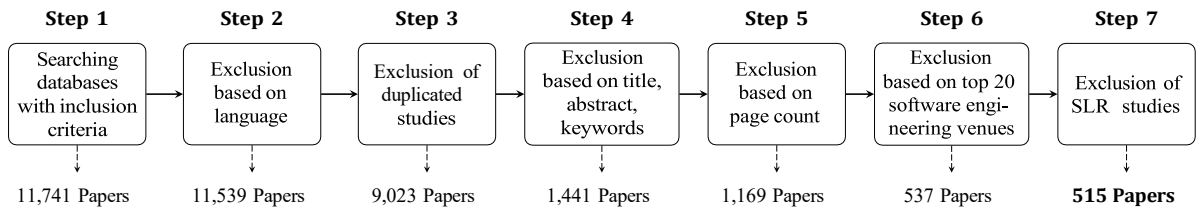

Figure 3: The systematic mapping process. From left to right we refined the papers we gathered to meet the scope of our project using Inclusion/Exclusion Criteria

resulting in a set of 515 papers from the original 11, 741.

Table 1: Selected publication venues from Google Scholar for exclusion Criteria V and the final number of papers studied from each venue.

| Publication | Paper count |
| --- | --- |
| ACM/IEEE Inter. Conference on Software Engineering | 63 |
| Journal of Systems and Software | 99 |
| IEEE Transactions on Software Engineering | 101 |
| Information and Software Technology | 99 |
| ACM SIGSOFT Inter. Symposium on Foundations of Software Engineering | 28 |
| Empirical Software Engineering | 32 |
| IEEE Software | 12 |
| ACM SIGPLAN Conference on Programming Language Design and Implementation (PLDI) | 1 |
| Mining Software Repositories (MSR) | 9 |
| IEEE/ACM Inter. Conference on Automated Software Engineering (ASE) | 49 |
| ACM SIGPLAN-SIGACT Symposium on Principles of Programming Languages (POPL) | 0 |
| Inter. Conference on Software Analysis, Evolution, and Reengineering (SANER) | 4 |
| Proceedings of the ACM on Programming Languages | 1 |
| Software & Systems Modeling | 3 |
| Inter. Symposium on Software Testing and Analysis | 3 |
| IEEE Inter. Conference on Software Maintenance and Evolution | 5 |
| Software: Practice and Experience | 4 |
| Conference on Tools and Algorithms for the Construction and Analysis of Systems (TACAS) | 0 |
| IEEE Inter. Requirements Engineering Conference | 2 |
| ACM SIGPLAN Symposium on Principles & Practice of Parallel Programming (PPOPP) | 0 |
| **Total** | 515 |

ACM Digital Library, IEEE Xplore, ScienceDirect, Scopus, and Inspec. We applied two broad *Inclusion Criteria* while performing a search in these libraries:

Inclusion Criteria I: Studies must have been published between 2000 and 2021,

Inclusion Criteria II: Studies must have been published in journals or conference proceedings.

Our initial search resulted in 11,741 Publications. To further narrow these down and following similar mapping studies in software engineering [75] we used a set of heuristic Exclusion Criteria to refine our search:

Exclusion Criteria I: Studies not presented in English.

Exclusion Criteria II: Duplicate studies,

Exclusion Criteria III: Studies not having variations of *Software* (SE, Program, Application, and App) and *Automation* (Automate, Automated, Automating, Automatization, Automatize) in title, abstract, and keywords,

Exclusion Criteria IV: Studies having less than six pages and hence are not considered as a full paper in the current software engineering proceedings,

Exclusion Criteria V: Studies not presented in the top-ranked venues according to Google Scholar [2],

Exclusion Criteria VI: Studies providing a systematic review of the literature, as they often discuss multiple approaches and themes and not aligned for a side by side comparison in **RQ3**.

We retrieved 515 Papers after applying the Exclusion Criteria distributed over the years, as shown in Figure 4. Appendix 1 shows a full list of these papers and their categorization, while We provide the details of the number of publications across different venues in Table 1.

### 3.2.2. Comparing State of the Art and Practice for Open Source Production Tools

To assess the current state of both art and practice, we meticulously analyzed 515 papers and aligned their findings with GitHub Categories. In the following sections, we delve into a comprehensive breakdown of our methodology.

**Data Synthesis:** For a side-by-side analysis of automation topics within the open-source ecosystem in both practical application and literature, we classified each paper according to the GitHub Marketplace categories (Table 2). Following the marketplace policies, the product owner must classify their automation tool into at least one category. Categories are common in marketplaces and offer end-users a convenient way to find products and services and possibly compare and choose among alterna-tives. Mobile app categories have proved to be a representative label for identifying the products' core functionality [47, 94]. Similarly, we used the GitHub Marketplace categories to identify the core scope of functionality perceived by the developers in our study.

**Mapping for a side-by-side comparison:** We manually categorized the papers based on GitHub Marketplace's categories. The first two co-authors of the paper carefully reviewed 515 paper abstracts and categorized them into up to three categories individually. In case of discrepancies, the Fourth co-author was provided with the list of debated papers and independently categorized them. The results were reviewed by the authors and finalized by the majority-based ruling. We calculated the degree of agreement among our annotators using Kappa's agreement coefficient (a measure of inter-rater reliability indicating how much of the observed agreement is due to chance) during the classification process [8]. The Kappa agreement above 0.6 is considered a substantial agreement [87]. In this case, the degree of Kappa's agreement was 0.71, indicating a substantial level of agreement among our annotators.

**Qualitative analysis of tools:** We also compared the perspective of researchers and practitioners on the relevance and connection between different categories, using both categories and descriptions. New technologies and innovations appear from the synergy between topics. For example, Vasilescu et al. [91] discussed approaches such as "related-by-usage" (two languages used by the same authors in the same projects) and "related-by-knowledge". Hence, we identified the most active and inactive intersection between the categories. In addition to comparing co-occurrence between categories, we compared the content and description of marketplace tools with state-of-the-art studies. We used short and full descriptions of tools to extract the key topics in the marketplace. In **RQ3** we extract these topics from the paper title and abstracts.

**Population and popularity:** Also, we compared the *population* of active researchers in automating a particular software process (category). We calculated the number of distinct App and Action developers and distinct paper authors in our data. We identified distinct GitHub developers using their unique IDs. For the authors of the papers, we performed a careful manual analysis of the authors' names and, if necessary, referred to their research profiles to distinguish between different authors who shared identical names. We compared the number of individuals active in each category in the marketplace (developers, **RQ1**) and in academia (researchers, **RQ3**).

Further, we compared the *popularity* of the studies with the products. As for the popularity in practice, we used the average number of installs (for Apps), and the average number of stars (for Actions) per each category as the proxy. We took the total number of installs or stars for all Apps or Actions in a category and divided it by the total number of products in that category. For academic popularity, we used Google Scholar's API [86] and calculated the average number of citations *per year* across all the papers mapped to each category. For each paper, we divided the number of citations by the number of years from the time they were published and calculated the average number of citations. We then ranked each category based on these popularity proxies, once for the marketplace and once for literature.

## 4. Empirical Results

In contrast to the known app stores for the research community, such as Google Play, the GitHub Marketplace provides a moderate degree of customization for attributes provided by the developers for their automation tools.

In section 3, we scraped attributes from 439 out of 440 Apps and 7,846 out of 7,878 Actions in addition to attributes of the platform itself. We divided these attributes as *Developer*, *Platform*, and *User* attributes and into *Characteristics*, *Features*, and *Policies* as defined in

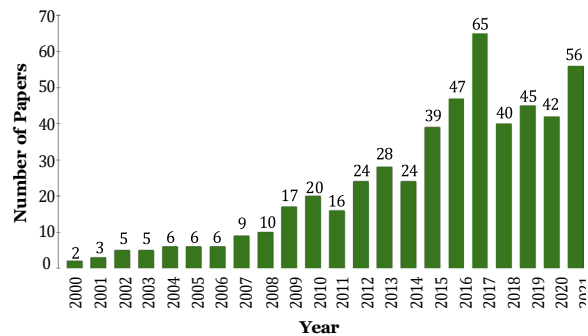

Figure 4: Number of categorized papers within our systematic mapping study (**RQ3**) in each year

Figure 1. We summarized our classification in Figure 2 and discuss our findings in detail in section 4.1 and 4.2

We retrieved a total of 462 relevant academic papers after applying the exclusion criteria noted in section 3.2 across the top 20 most popular and relevant publications. Table 1 contains the details of these publications and the count of papers from each. Appendix 1 shows a full list of these papers.

*4.1. Marketplace Characteristics (RQ1)*

Characteristics of a marketplace are entities that are not under the direct influence of the marketplace owner [35]. In this section, we discuss our findings regarding the Characteristic component aligned with the conceptual model depicted in Figure 1. We discovered nine Characteristics in GitHub Marketplace across three attribute categories. These attributes are depicted in Figure 2. The Developer attributes that are not directly controlled by GitHub Marketplace are *Name, Categories, Short Description, Full Description, and Releases*. The User attributes of such Characteristics include *Contributors, Number of Stars, and Number of Installs* while the Platform attributes contain only *Developer*. In the following subsections, we will briefly define each of these attributes while presenting our findings.

Table 2: Categories of Apps and Actions and their definition as defined by GitHub Marketplace.

| ID | Category | Category Description |
| --- | --- | --- |
| C1 | API management | Structure your API infrastructure to enable various internet gateways to interact with your service. |
| C2 | Testing | Eliminate bugs and ship with more confidence by adding these tools to your workflow. |
| C3 | Utilities | Auxiliary tools to enhance your experience on GitHub. |
| C4 | Reporting | Get insights into how your teams are developing software using GitHub. |
| C5 | Continuous integration | Automatically build and test your code as you push it to GitHub, preventing bugs from being deployed to production. |
| C6 | Publishing | Get your site ready for production so you can get the word out. |
| C7 | Support | Get your team and customers the help they need. |
| C8 | Project management | Organize, manage, and track your project with tools that build on top of issues and pull requests. |
| C9 | Code review | Ensure your code meets quality standards and ship with confidence. |
| C10 | Deployment | Streamline your code deployment so you can focus on your product. |
| C11 | Chat | Bring GitHub into your conversations. |
| C12 | Community | Tools for the community (Action only) |
| C13 | Container CI | Continuous integration for container applications. |
| C14 | Dependency Mgmt. | Secure and manage your third-party dependencies. |
| C15 | AI Assisted | Tools that are super-powered with AI (artificial intelligence) to help you be a better developer. |
| C16 | Open Source Mgmt. | Running open-source projects can be hard. Here are some tools to make that process a little more fun and a ton more manageable. |
| C17 | Security | Find, fix, and prevent security vulnerabilities before they can be exploited. |
| C18 | Monitoring | Monitor the impact of your code changes. Measure performance, track errors, and analyze your application. |
| C19 | Code quality | Automate your code review with style, quality, security, and test-coverage checks when you need them. |
| C20 | Localization | Extend your software's reach. Localize and translate continuously from GitHub. |
| C21 | Desktop tools | Developer tools that are run natively on your local machine. |
| C22 | Mobile | Improve your workflow for the small screen. |
| C23 | IDEs | Find the right interface to build, debug, and deploy your source code. |
| C24 | Mobile CI | Continuous integration for Mobile applications. |
| C25 | Code search | Query, index, or hash the semantics of your source code. |
| C26 | Code Scanning Ready | Static analysis, dynamic analysis, container scanning, linting, and fuzzing tools that integrate with GitHub Code Scanning SARIF Upload. |
| C27 | Learning | Get the skills you need to level up. |
| C28 | Time tracking | Track your progress, and predict how long a task will take based on your coding activity. |
| C29 | Game CI | Tools for building a CI pipeline for game development (Action only). |
| C30 | Backup Utilities | Utilities providing periodic backups of your GitHub data. |
| C31 | Content Att. API | Tools that use the Content Attachments API (App only). |
| C32 | GitHub Created | Tools created by the team at GitHub with love (App only). |

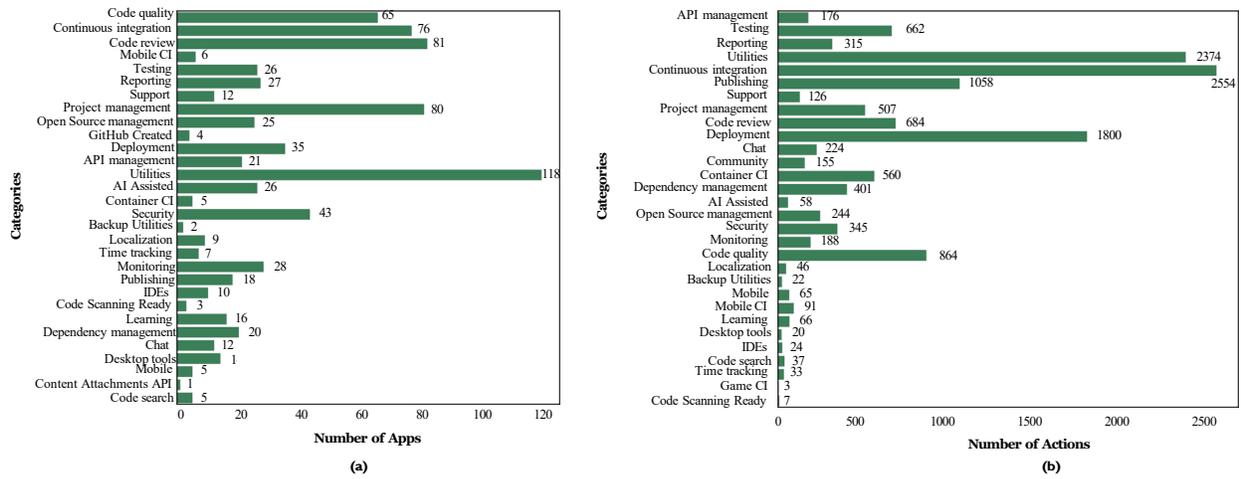

Figure 5: Each App or Action is listed in one or more categories by its developer. (a) shows the categories and frequencies for Apps, and (b) shows the categories and frequencies for Actions.

*4.1.1. Name of automation tools*

Each automation tool on GitHub Marketplace requires a Name attribute to act as a unique identifier of the tool. The name is limited to 255 characters and cannot match an existing GitHub-hosted account/tool (unless it is the developer's own user or organization name). Additionally, for the GitHub Actions, the name cannot be the same as pre-defined GitHub Marketplace categories. Even though the marketplace owner has some regulations when selecting a name, developers have complete freedom to select any name that fits their product.

*4.1.2. Categories of automation tools*

Each tool on the marketplace can be listed in up to two categories (primary and secondary) by its developer. Currently, 32 categories exist in the marketplace. These categories and their definition by the GitHub Marketplace are provided in Table 2. There are two categories exclusive to Apps, and two are exclusive to Actions. Even though, the categories are pre-defined by GitHub Marketplace, it is the developers who decide on which categories to select for their product, making this attribute a Characteristic.

Figure 5 shows the number of Apps and Actions in each category. Overall, we have extracted data from 8,318 automation tools for the purpose of our study. We have observed that Apps sometimes belong to more than two categories. 39 Apps have three categories, and two Apps have four categories (regardless of marketplace rules on a limit of two categories for each product). For Apps, the categories 'Utilities' (118 Apps), 'Code Review' (81), and 'Project Management' (80) have the highest population. 'Continuous Integration' (2,554 Actions), 'Utilities' (2,374), and 'Deployment' (1,800) are the categories with the most number of Actions.

> *'Utilities' include the most number of Apps while 'Continuous Integration' has the most number of Actions in the marketplace.*

*4.1.3. Short and Full description of automation tools*

GitHub allows product owners to add brief descriptions for their products in the marketplace to explain the product's main functionality. 18.64% of the Apps (82 out of 440) have short descriptions of less than 40 Characters while 33.64% of the Apps (148 out of 440) have short descriptions that exceed 80 Characters. For Actions, 28.90% and 15.65% have a short description length of below 40 Characters and above 80 Characters, respectively. Figure 6 shows the distribution of the number of characters for the short description of Apps and Actions.

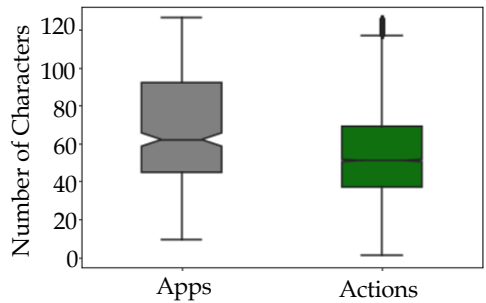

Figure 6: Length distribution of short descriptions for automation tools on the GitHub Marketplace

Each product has a dedicated page and URL on the marketplace that includes an elaborate description of the app or action. For Actions, GitHub does not provide guidelines regarding the format or length of this description. However, for Apps, this full description should consist of two parts: a required 'Introductory description', and an optional 'Detailed description' [27].

The 'Introductory description' is featured on the App's landing page top. GitHub suggests a concise high-level summary of 150-250 characters. Additional details go in the 'Detailed Description,' where developers present three to five value propositions, each in two sentences. GitHub advises a title and a paragraph for each proposition, avoiding complete sentences in titles. Detailed descriptions can use up to 1,000 characters. Developers can include up to five screenshots on the landing page by following GitHub's guidelines for screenshots [28].

We mined the full descriptions of the tools to understand the features offered by the Apps and Actions in each category. Similar to the work of Al-Subaihin et al. [5] on mining features from mobile App descriptions, we refer to a feature as a claimed functionality mined from the product description. Table 3 shows the mined features in each category along with their term frequencies. The most frequent feature of the products in the GitHub Marketplace is issue management. Categories 'Project management' (127 Mentions of this feature), 'Open Source management' (54), and 'Utilities' (53) all have products that offer a variety of these features, such as open or close issues, label issues, find an issue, or comment on an issue using a bot. Among these, issue labeling is the most prevalent feature. Unit test also appears frequently in products from the categories 'Testing' (103), and 'Reporting' (50). Also, Apps and Actions in the marketplace that facilitate static analysis mainly fall in 'Code quality' (72), 'and Security' (21). The categories 'Dependency management' (17), 'AI-Assisted' (11), 'Utilities' (11), and 'Open Source management' (6) all have products that help with managing dependency. These products mainly aid developers in updating dependencies, resolving dependency conflicts, creating dependency graphs, and documentation. A variety of products in overlapping categories also help developers with code coverage, API documentation, and automating the deployment.

### 4.1.4. Developers of automation tools

The marketplace also contains information about the developers who have published their tools on the platform. Overall, 5,928 developers have been sharing and distributing their automation tool on GitHub Marketplace by the time of this study. The majority of these developers solely share GitHub Actions (92.93%), while 6.21% focus on distributing GitHub Apps exclusively. Only a small number of developers published both Actions and Apps (less than 1%). Additionally, 20.70% (1,227) of the developers have more than one product in the marketplace, and on average, each developer has 1.40 automation tools in the marketplace. Among these developers *Azure* has the most number of products (one App and 35 Actions). This developer has the highest number of Actions by one provider in the GitHub Marketplace. *Devbotsxyz* owns the highest number of Apps with five Apps (only one Action).

### 4.1.5. Releases of Actions

For each Action, the GitHub Marketplace contains data on the various releases of the Action, allowing the developers seeking to use the tool to choose a different version of the tool if applicable. 42.64% of the Actions only have one release, while 1.43% of the Actions have nine releases or more. Additionally, we used the Pearson correlation coefficient (r) between the number of releases and the number of stars to discover that there is a weak correlation between the two (r= 0.19).

### 4.1.6. Contributors to Actions

The marketplace lists all the contributors to an Action. Contributors are the software developers who open an is-sue, propose a pull request, or commit any change to the default branch of an Action repository. 7,111 Actions (90.63%) in the marketplace have at least one contributor. 61.81% of Actions have a single contributor. In contrast, a total of 145 Actions have 12 Contributors, which is the largest number of contributors for an Action on GitHub Marketplace.

Table 3: Top features in each marketplace category (extracted from the long description of products). The numbers in parenthesis stand for the frequencies of each feature appearing in the product descriptions.

| Category | Top features |
| --- | --- |
| AI Assisted | label issue (21), unit test (12), dependency conflict (11), search code (4), code enhance (3) |
| API management | api specification (15), api documentation (13), swagger ui (13), performance test (5), schema change (3) |
| Backup Utilities | backup repository (5), data retention (4), daily backup (3), run workflow (3), cloud storage (2) |
| Chat | send message (110), slack notification (76), custom message (34), discord webhook (30), slack bot (29) |
| Code quality | code coverage (113), run test (82), static analysis (72), code review (27), automate code (5) |
| Code review | linting process (64), code coverage (59), automatically merge (6), test coverage (5), automate code (5) |
| Code scanning ready | infrastructure code (4), code scan (4), check vulnerability (4), automatically merge (2), automate deployment (2) |
| Code search | issue create (7), unit test (5), code enhance (3), best practice (3), search code (3) |
| Community | api key (25), push branch (20), create issue (17), open source (17), issue comment (14) |
| Container CI | container registry (126), run docker (82), build image (74), push image (61), machine learning (3) |
| Content Attachments API | real time (3), account login (2), embed issue (2), team collaboration (2), collaboration leanboard (2) |
| Continuous integration | docker image (347), run test (320), run workflow (206), docker container (160), unit test (36) |
| Dependency management | run test (54), project documentation (31), dependency conflict (11), fix vulnerability (7), update dependency (6) |
| Deployment | docker image (226), run build (136), cloud deploy (4), easy build (3), automatic deployment (3) |
| Desktop tools | cake script (12), UI test (5), desktop app (5), quick filter (5), issue tracker (2) |
| Game CI | resource pack (11), optimize resource (5), build renpy (2), generate zip (2), file distribution (2) |
| GitHub Created | jira issue (4), friendly bot (3), learn skill (3), post comment (2), connect jira (2) |
| IDEs | code editor (4), smart IDE (3), code deploy (2), app inspection (2), sass solution (2) |
| Learning | update readme (8), execute workflow (7), smart IDE (3), learn skill (3), share knowledge (3) |
| Localization | check spelling (5), generate translation (4), machine translation (4), google translate (4), continuous localization (2) |
| Mobile | run test (26), android emulator (9), unit test (9), android app (8), deploy apps (3) |
| Mobile CI | run android (14), android emulator (8), android CI (7), upload artifact (6), deploy apps (3) |
| Monitoring | send notification (19), run lighthouse (16), html report (14), graphql inspector (5), crash reporting (3) |
| Open Source management | create workflow (31), close issue (25), label issue (29), post comment (5), merge pull-request (3) |
| Project management | create issue (68), label issue (59), create release (45), kanban board (9), track progress (7) |
| Publishing | create release (266), release note (178), docker image (132), build deploy (102), release tag (94) |
| Reporting | send message (52), unit test (50), slack notification (46), test report (33), see dashboard (2) |
| Security | security scan (53), scan repositories (51), security analysis (36), fix vulnerability (25), static analysis (21) |
| Support | support request (14), development support (13), label issue (10), powerful analytics (3), customer service (3) |
| Testing | unit test (103), test playbook (88), code coverage (66), test report (47), test automation (6) |
| Time tracking | todoist api (10), update readme (9), todo list (8), task organization (2), productivity growth (2) |
| Utilities | project documentation (159), docker image (128), create workflow (108), dependency conflict (11), pull-request review (8) |

*4.1.7. Number of Stars for Actions*

To keep track of favourite projects in addition to showing appreciation to the Action developers, the GitHub user can use the Star system on the relevant repositories [25]. The marketplace displays the star count of Actions and may recommend related content to users based on their starred repositories on their dashboards. Furthermore, many of GitHub's repository rankings depend on the number of stars a repository has. Our current analysis used the number of stars as a popularity metric for the Actions on the marketplace. Among various Actions on the marketplace, Super-Linter developed by GitHub has the highest number of stars for Actions with 6,800 Stars. This is followed by yq - portable YAML processor developed by mikefarah with 3,900 Stars. Figure 7-(b) demonstrates the distribution of the star ratings among Actions. Considering the star distribution over the number of existing Actions in the Marketplace categories, 'Time tracking' and 'Community' were the top two popular categories. In contrast, 'Game CI' and 'Backup Utilities' were the least popular ones.

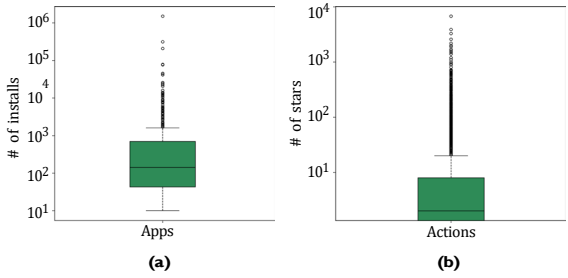

Figure 7: Distribution of number of installs for **(a)** Apps and number of stars for **(b)** Actions on the marketplace as a measure of popularity.

*4.1.8. Number of Installs of Apps*

Due to GitHub's mechanism of masking the number of installs for the 'Recommended Apps', we based our analysis on 85% of the Apps (374 Apps out of 440) for which the number of installs is provided. On average, Apps have been installed 7,530.12 times from the GitHub Marketplace. However, 92.25% of Apps have less than the average number of installs indicating that there are highly popular Apps that users favour over others. As for categories, the highest average number of installs belongs to 'GitHub Created' (with 594,500 Installs), 'Learning' (115,748.08), and 'API Management' (75,320). The overall highest number of installs was for GitHub Learning Lab with 1.5 Million installs, followed by Travis CI with 314,000 Installs. Slack + GitHub is the third most installed App with 207,000 Installs. Figure 7-(a) is the boxplot distribution of the 'number of installs' for Apps.

> *The popularity of Apps and Actions can be measured based on the number of installs and the number of stars, respectively. The Apps created by GitHub (C32) and the Apps on 'Learning' (C27) are the top two App categories with the highest average number of installs. 'Time tracking' (C28) and 'Community' (C12) have the highest average number of stars for Actions.*

*4.2. Marketplace Features and Policies (RQ2)*

As proposed by Jansen and Bloemendal [35], the Features and Policies, together, are the components of a marketplace that are under the direct influence of the marketplace owner (in this case, GitHub Marketplace). Furthermore, Al-Subaihin et al. [4] have explored the impact of platforms on software engineering practices. While their study is only limited to mobile app development, the results demonstrate the impact of platforms on software is non-trivial. In this section, we will discuss the Features & Policies discovered in GitHub Marketplace aligned with the model illustrated in Figure 1. GitHub Marketplace, as a platform owner, directly influences six attributes across two classes: Developer and Platform attributes. These attributes are Developer Link, Supported Languages, Verified Domains, Pricing, Recommended, and GitHub Verification, with the first four belonging to the former category and the last two belonging to the latter category. In the following sections, we briefly explain the definition of each attribute and summarize the existing status quo by the date of our study.

*4.2.1. Developer Link for Apps*

The App listing page on GitHub Marketplace requires two mandatory links for customer support (redirecting clients to a web page for getting technical help, product information, or account information) and the App's privacy policy statement [26]. Additionally, the marketplace allows developers to include optional links such as company, status, and documentation URLs. The company URL is a link to the company's website, while the status URL is a link to a web page that shows the App's status. Current and historical incident reports, web application uptime status, and scheduled maintenance can all be seen on status pages. The documentation URL leads to documentation that demonstrates how to use the App.

*4.2.2. Programming Language for Apps*

The GitHub Marketplace has a pre-defined set of programming languages from which developers can select up to ten languages supported by their product. This attribute is optional, and the selected languages are displayed on the App's listing page. Developers are limited to a selection of programming languages defined by the GitHub Marketplace.

We identified 80 distinct programming languages that Apps support on the marketplace. However, 234 Apps (53.30%) did not document their supported languages. Among the remaining Apps, JavaScript with 71.22% (146 out of 205Apps), Python with 59.51% (122 out of 205), Java with 57.56% (118 out of 205), Ruby with 46.83% (96 out of 205), and Go with 42.93% (88 out of 205) are the top five languages supported by the Apps.

We also studied the co-occurrence of programming languages among the Apps [91]. Most frequently, JavaScript and Python appear together in 107 Apps, followed by JavaScript and Java, which were jointly supported in 106 Apps. The Apps supporting automation of the workflow for JavaScript programs are also likely to support programs in Java, Python, or both. 77 out of 80 languages are supported by Apps that support at least one other language. ABAP, LookML, and OCaml are the only languages not associated with any other languages.

*4.2.3. Policies of Verified Domains for Apps*

For the App developers to offer paid plans for their product and to have marketplace badges on their App listing, they are required to complete the publisher's verification process for their (or their organization's) account. As part of this process, the publisher must ensure that their organization has verified ownership of their domain. 43.51% (191 out of 439) Apps on the marketplace have verified domains.

*4.2.4. Pricing Policies for Apps*

GitHub Actions are free for both GitHub-hosted runners and self-hosted runners. For self-hosted runners, each GitHub account receives a certain amount of free minutes and storage. For example, *GitHub Free* has 500 MB of storage and 2,000 Minutes/Month, compared to 1 GB storage and 3,000 Minutes/Month for *GitHub Pro* [23]. Customers are billed for additional usage of Actions beyond the storage or minutes included in their account.

For Apps, GitHub Marketplace pricing plans can be *free*, *flat rate*, or *per-unit*. *Free plans* are completely free for users and developers are encouraged (not enforced) to offer free plans as a way to promote open-source services. There are two types of paid pricing plans- *flat rate* and *per-unit*. Currently, there are 223 different pricing plan names on the marketplace, 51 of which are variations of *Free*, for example, Free, Open source, Basic, Default, Starter, Hobby, Just enjoy it, etc. On the marketplace, 96.81% (425) Apps have a free tier compared to 19.36% (85) Apps with at least one paid plan. 80.64% (354) Apps have only the free plan in contrast to 3.19% (14) Apps that have only paid plans. Finally, 16.17% (71) Apps have both free and paid plans.

*4.2.5. Recommended Apps*

15.0% (66 out of 440) of the Apps on the GitHub Marketplace have a label stating *Recommended*. The number of installs for these Apps is not provided, and GitHub Marketplace does not provide any official description of the eligibility criteria for this tag. We checked the list of recommended Apps across different users as well as for three users over time. The set of recommended Apps is static and is not curated based on a user or usage over time. Further, we hypothesized that this could be relevant to App verifiability and authentications. However, our observations showed that these recommended Apps have a lower percentage (18.18%) of verification. Only 22.72% of these recommended Apps have a verified domain compared to 47.06% of the non-recommended Apps with at least one verified domain. 'Localization' (33.33%), 'Time Tracking' (28.57%), and 'Support' (25.0%) are the categories with the highest proportion of recommended Apps.

*4.2.6. GitHub Verification for All automation tools*

GitHub Marketplace verifies creators and products separately. To offer paid plans for GitHub Apps or to include a marketplace badge in the App listing, the App developer must complete the publisher's verification process for their organization and become a 'verified creator'. As part of this process, the publisher must ensure that their basic profile information is accurate, including an email address for support and updates from GitHub. GitHub reviews the details and informs the organization once their publisher verification is complete. Once the publisher (and/or organization) has been verified, they can publish paid plans for their Apps. Actions with the *verified creator* badge indicate that GitHub has verified the creator of the Action as a partner organization.

Two-factor authentication also needs to be enabled for the organization. The organization must have verified ownership of their domain and ensure that a 'Verified' badge displays on the organization's profile page. However, there are two possible levels of verification for Apps:

· App meets the requirements for listing: These Apps meet the listing requirements but the publisher has not been verified. These Apps cannot change their pricing plan until the publisher successfully applies for verification.

- Publisher domain and email verified: These Apps are owned by an organization that has verified ownership of their domain, confirmed their email address, and required two-factor authentication for their organization. In other words, the App went through the publisher verification process and was successfully granted a 'verified creator' status.

Our results indicate that overall there are 3.67% (304 out of 8,225) Verified Creators in the marketplace. 2.67% (209 out of 7,846) of the Actions on the marketplace have verified creators. 69.09% (304 out of 440) of the Apps in the marketplace are *not* verified by GitHub Marketplace and do not have any level of verification badge. Among the 136 Verified Apps, 41 have the 'App meets the requirements for listing' status, while 95 hold the 'Publisher domain and email verified' status.

> *Majority (79.3%) of the developers have published only one tool on the marketplace. GitHub policies require developers to verify their organization so they can publish paid plans for their Apps. Only 2.67% Action developers and 21.59% App developers are verified creators on the marketplace. Despite flexible pricing plans for the tools on the marketplace, only verified creators can distribute paid tools.*

### 4.3. Research vs. practice of open source Production automation Tools (RQ3)

Innovation thrives when synergies are identified and extended. The symbiotic relationship between academia and industry serves as a crucial catalyst for innovation, especially within the context of open source, which is known for its success in open innovation [92, 51]. Automated software production has attracted significant attention among researchers. We systematically mapped 515 papers to compare side by side with the tools on the GitHub Marketplace.

#### 4.3.1. Comparing the Categories and Nature of Open Source Production Tools

Following the systemic procedure described earlier (see Section 3.2), two authors independently mapped the papers included in the study into the automation tool categories (32 Categories overall). The detailed mapping of each paper to each category is provided in Appendix 2. Figure 8 shows the heatmap of the number of papers per year in each category. The category of 'Code quality' stands out as the foremost area of focus, comprising 140 out of the total 462 papers. Examining the annual distribution, we noted that the highest number of papers dedicated to discussing automation tools is in 'Code quality' (18) in 2017, followed by the papers in 'Code quality' and 'AI-Assisted' (15 each) in 2021.

To understand the contents of these papers in further detail, we extracted bi-grams from the titles and abstracts of all the papers in each category. We followed the same protocol and process for extracting bi-grams for tools on the marketplace. These bi-grams represent the key features or phrases in each category.

In the literature, we found that several common features appear across multiple categories. One such feature is bug reporting, which is found in 11 categories, namely: 'Project management' (46 mentions), 'AI Assisted' (45), 'Code quality' (30), 'Code review' (27), 'Reporting' (22), 'Support' (20), 'Code search' (18), and 'Monitoring' (12). Another widely shared feature is requirement traceability, which is the focus of papers in several categories. Specifically, 'Code review' (21), 'Utilities' (13), 'Dependency management' (11), 'IDEs' (10), 'AI Assisted' (10), and 'Localization' (8) all have papers discussing the topic of requirement traceability. Additionally, several categories share papers facilitating static analysis. These categories include 'Code quality' (14), 'Code Scanning ready' (10), 'Utilities' (6), and 'Deployment' (4), all of which have papers related to this topic. Lastly, the topic of test automation is addressed in papers falling under the categories 'Testing' (17), 'Mobile' (4), 'Mobile CI' (4), and 'Open source management' (2). These common features and their presence across multiple categories highlight their significance and relevance in the literature.

> *Studies in different categories often discuss similar topics/features. The most common feature in literature is 'bug reporting' (appearing in 11 categories), followed by 'requirement traceability' (six categories), 'static analysis' (four categories), and 'test automation' (four categories).*

We also looked into the number of distinct authors per category. There are 1,434 distinct authors for 462 papers in our study. 'Code quality' with 401 authors, 'Testing' with 331 authors, and 'AI Assisted' with 297 authors are the top three categories that have the highest number of active researchers. These are also the top three categories with the most papers. 25.54% of the papers (118 out of 462) have four authors, which is also the median number of authors per paper.

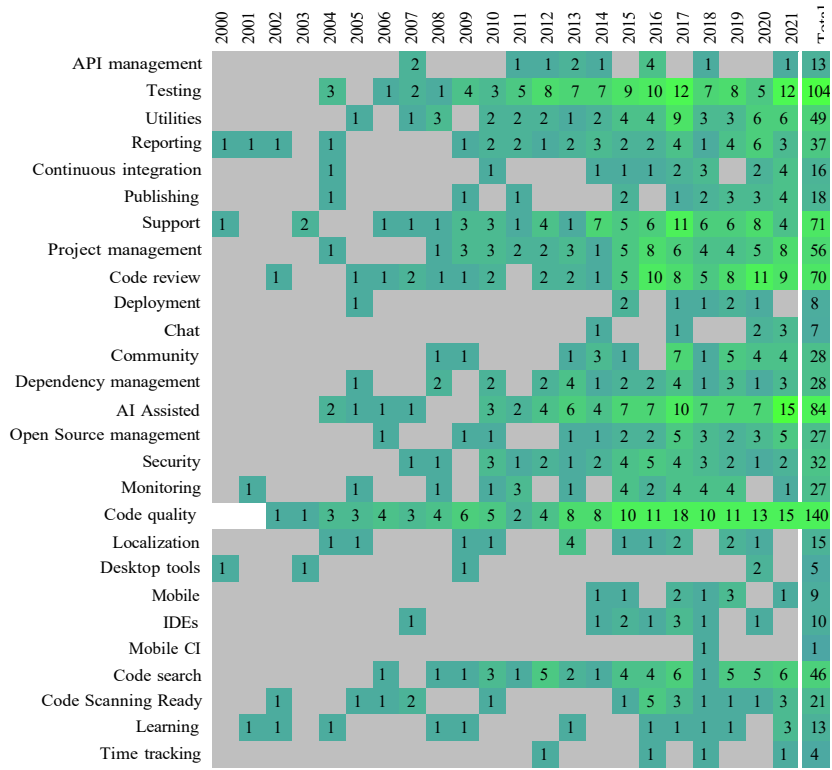

Figure 8: Number of papers in each year per category. Categories with zero number of papers were excluded from the heatmap.

'Publishing' tools help developers with their software releases. While the marketplace offers tools that facilitate release creation (266), generating release-notes (178), and release tags (94), the literature discusses more efficient release scheduling (4). Both products and papers offer tools for cloud deployment. The set of features covered by the 'Mobile CI' category in the marketplace were divergent from their counterparts in literature. Tools on the marketplace focus on automating the build, test, release, monitoring, and deployment of projects; however, we only mapped one paper in this category, which discusses automated tests for mobile apps.

'Code quality' includes that help with static analysis. As for the 'Code review' category, the marketplace has products for automatically merging pull requests (6), test coverage (5), code linting (64), and a variety of other use-cases. Literature, on the other hand, discusses code analysis (14), program comprehension (14), bug reporting (27), etc in this category. For 'API Management', both literature and marketplace products assist with API documentation and API specification. The marketplace offers products under 'Dependency management' that aid developers in finding and resolving dependency conflict (11) as well as updating dependency (6). Similarly, literature in this category proposes tools that help with co-changing dependency (4).

'Game CI', 'Container CI', and 'Continuous integration' were also disparate from the literature in terms of features. 'Game CI' contained tools for 'building CI pipeline for game development [1] such as integrating Unity build engine commands, offer notifications on the new releases, or set up static data products that host game assets. Tools in the 'Container CI' category of the marketplace assist in creating, retrieving, and registering container images across multiple platforms in addition to automating tasks such as pushing and running builds for platforms such as Apache Kafka. Tools in 'Continuous integration' offer integration of different tools such as Maven CLI, Armvirt, and Pipenv into GitHub repositories to facilitate developers' activities. We did not find any paper directly related to either 'Game CI', 'Container CI', or 'Continuous integration' within our systematically retrieved list of studies.

*4.3.2. Comparing Population and Popularity*

We chose three approximations for measuring and comparing the popularity of particular information tasks between academia and practice;

**First,** we compared the number of publications in each category with the number of tools offered in those categories (See Figure 9-b),

**Second,** for each category, we compared the percentage of developers who published tools on GitHub against the percentage of researchers who published papers in that category (See Figure 9-a),

**Third,** we compared the average number of installs and stars (for GitHub tools) with the average number of citations per year (for papers) for each category (See Figure 10).

First, we calculated the percentage of the papers and the marketplace products (Apps and Actions) in each category. The difference is less than one Percentage Point (pp) for nine categories and more than 10 pp for seven other categories. The distribution of papers and marketplace products varies greatly across categories. For example, 'Continuous Integration' is the #1 and #4 category for Actions and Apps, respectively. However, this category contains only 16 (3.46%) papers and is ranked 17 out of 27 Categories with at least one paper. On the other hand, 22.51% (104) Papers in our mapping study fall under the 'Testing' category, while only 5.92% (26) Apps and 8.44% (662) Actions on the marketplace fall under this category. 10 categories have a higher percentage of marketplace products. This difference is more than 15 pp for the categories 'Continuous integration' (28.28 pp), 'Deployment' (20.42 pp), and 'Utilities' (19.64 pp).

In contrast, The categories of Code quality' (19.09 pp), AI assisted' (17.17 pp), Testing' (14.21 pp), Support' (13.7 pp), along with 17 other categories, exhibit a higher percentage of research papers mapped to these categories compared to the percentage of marketplace products. There were four categories on the GitHub Marketplace for which we did not find any papers within our systematic mapping study, namely 'Container CI', 'Game CI', 'Backup Utilities', and 'Content Attachments API'. Barring these four categories, 'Mobile', 'Time tracking', 'API Management', 'Desktop tools', 'Mobile CI', and 'Dependency management' had the least percentage point difference (ranging from 0.2 pp to 0.98 pp) between the proportion of the papers and the marketplace products. Figure 9(b) shows the percentage of the number of papers in the marketplace versus the systematic mapping.

Second, we looked into the distinct number of developers. There are 1,434 distinct authors and 5,937 distinct developers in our dataset of 462 Papers and 8,318 GitHub products (440 Apps and 7,878 Actions). We presented the number of developers (Figure 11) and the number of authors per category in **RQ1** and **RQ3** (See Figure 9-a). By comparing the two, we found that researchers and practitioners are interested in different subject categories. Figure 9 shows the percentage point difference between the contribution of researchers and practitioners for each category. For eight categories, there is more than 10 Percentage Point (PP) difference between the percentage of researchers and practitioners.

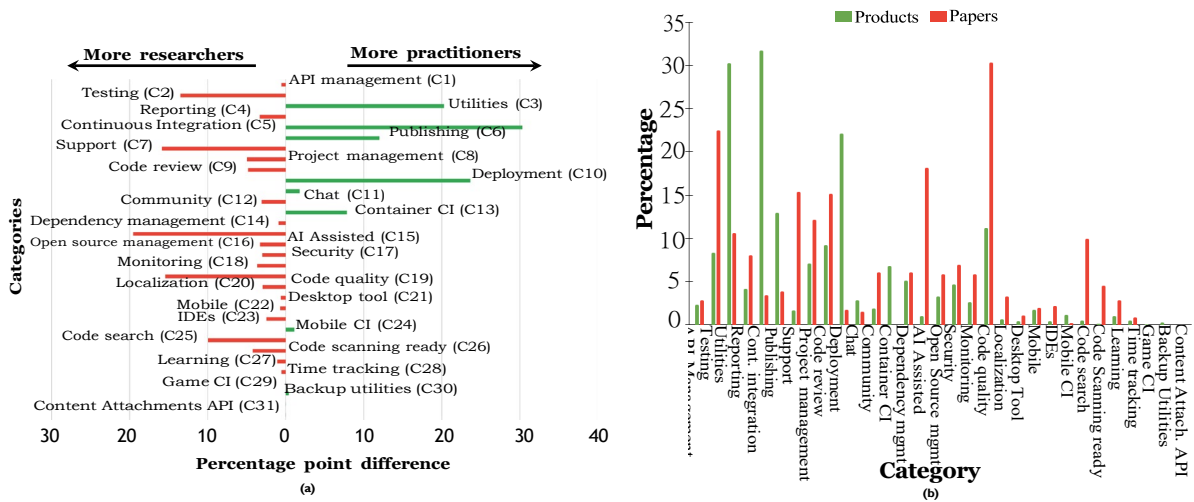

Figure 9: Difference in (a) the number of practitioners and the number of researchers (red shows a higher percentage of researchers working in a category, while blue indicates a higher percentage of practitioners); and (b) the percentage of products (blue) and papers (red) in each category.

> *There is a considerable difference in the distribution of papers and marketplace products across categories. 'Continuous Integration' and 'Code quality' are notable examples, with the former having a high percentage of marketplace products (pp = 28.28) and the latter having a high percentage of mapped papers (pp = 19.09).*

20.71% (297 out of 1,434) Researchers are publishing in the 'AI-Assisted' category. In contrast, only 1.15% (68) Developers are building Apps and Actions in this category (this implies a 19.56 difference in the Percentage Point (pp)). We have similarly striking gaps between the percentage of authors and developers working in the 'Support' (15.86 pp more authors) and 'Code quality' (15.43 pp more authors) categories.

> *In the open-source community, there exists a notable divergence between the research and practice domains. Researchers predominantly concentrate their efforts on enhancing 'Code quality,' whereas practitioners exhibit a strong inclination towards 'Continuous integration.' Remarkably, we observed a significant discrepancy of more than 10 percentage points in active participation between researchers and practitioners in a quarter of the categories (eight out of 32).*

21 Categories out of 32 have a higher percentage (ranging from 0.53 pp to 19.56 pp higher) of authors compared to the percentage of developers working in those categories. On the other hand, only 3.91% (56) of the authors are publishing in the 'Continuous integration' category compared to 34.23% (2,032) of the developers contributing to the marketplace in this category. Glaring gaps such as this, where a higher percentage of developers are involved compared to researchers, also exist in other categories such as 'Deployment' (23.67 pp more developers) and 'Utilities' (20.3 pp more developers). Categories with the least difference between the proportion of researchers and developers are 'API Management' and 'Time tracking' (both having 0.53 pp more researchers), 'Desktop tools' (0.63 pp more researchers), 'Mobile' (0.72 pp more researchers), and 'Mobile CI' (1.1 pp more developers).

While developers are the providers and publishers of the Actions, "contributors" can make changes to these open-source projects as well. Besides having a developer, most Actions (90.63%) in the marketplace have one or more contributors. 64.49% of these contributors are working in the 'Continuous integration' category compared to 3.91% of the authors. Categories 'Utilities' (41.37 pp), 'Deployment' (39.75 pp), and 'Publishing' (19.75 pp) have considerably higher percentages of con-tributors compared to authors. On the other hand, 'AI-Assisted' (21.85 pp), 'Support' (17.49 pp), and 'Code search' (10.63 pp) have a higher percentage of authors than contributors on the marketplace.

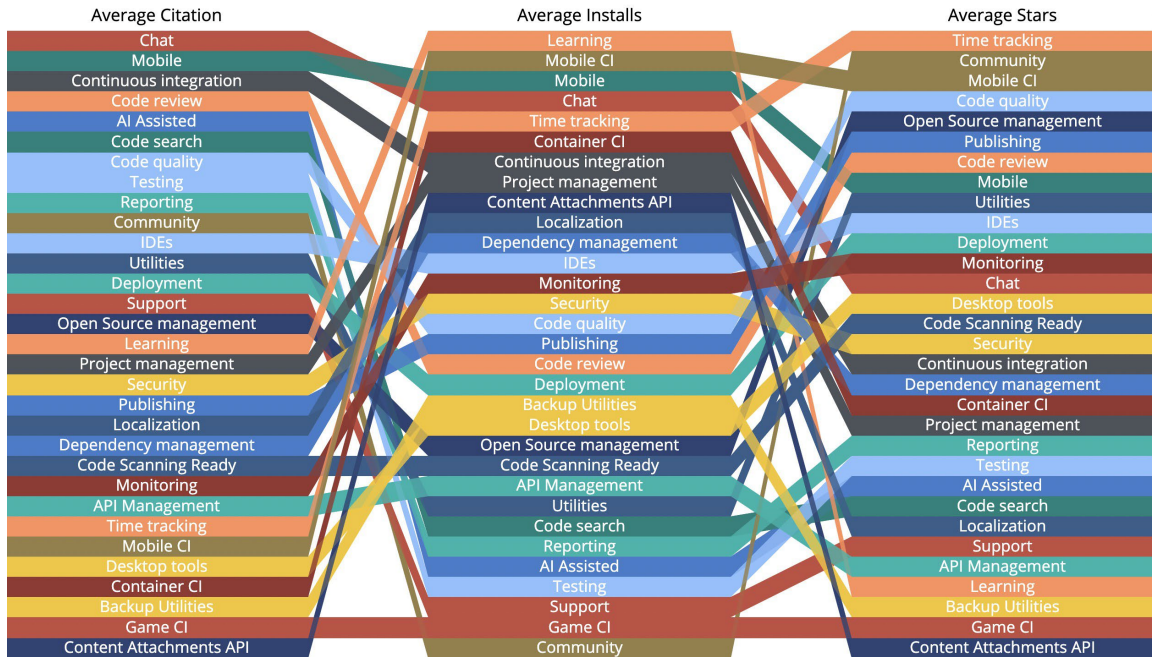

Figure 10: Category rankings are based on three factors: the average number of paper citations per year (left), the average number of App installs (middle), and the average number of Action stars (right).

Third, we compared the usage and popularity of these topics. Figure 10 shows a bump chart that compares the rankings of categories based on the average number of paper citations, App installs, and Action stars. The more popular categories in each column are shown on top, while the least popular categories are on the bottom. The category 'Mobile CI' displays the most striking gap in popularity between researchers and practitioners. It is ranked $2^{nd}$ and $3^{rd}$ in terms of average installs and stars, respectively, while appearing at the $26^{th}$ position in the average citation ranking (one paper compared with 91 Actions in this category). 'Time tracking' is another category that is significantly less popular among researchers (#25) compared to practitioners (#1 among Action users and #5 among the App users). 'Container CI,' 'Desktop tools,' 'Monitoring,' and 'Dependency management' are other categories that are more favored by practitioners when compared to researchers. On the other hand, the categories 'Code review', 'AI-Assisted', 'Code search', 'Testing', and 'Support' are more popular among researchers than both App and Action users. The category 'Game CI' ranks #30 (out of 31) in all three metrics, making it one of the least popular categories both in research and in practice. 'Content Attachments API' (#31) and 'Backup Utilities' (#29) are also unpopular among researchers and Action users although the categories rank $9^{th}$ and $19^{th}$, respectively, in terms of the average number of installs.

> *Among the popular categories in the Marketplace, 'Mobile CI' exhibits the largest disparity in citations (popularity of use) from researchers. On the contrary, 'AI Assisted' papers receive significant citations from researchers, displaying the most substantial contrast when compared to their popularity in the marketplace. 'Game CI' is equally overlooked and under-referenced by both researchers and practitioners.*

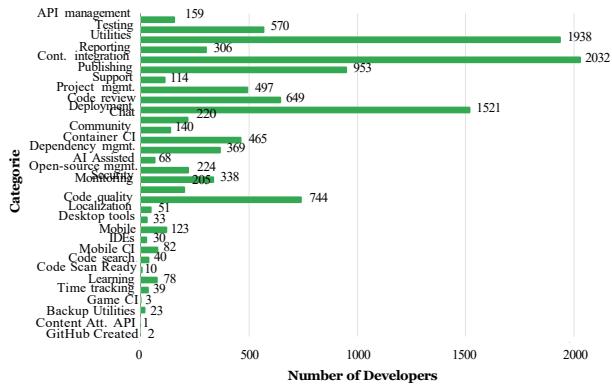

Figure 11: Number of developers (Apps and Actions) per category

## 5. Discussion and Implications

This paper involves three main contributions.

**First,** we performed the very first study on analyzing the GitHub Marketplace. The study of marketplaces has been of interest in the software engineering community since the emergence of mobile app stores. While there have been intense ongoing research and case studies in the domain of open-source software engineering, the marketplace, which is the platform for officially sharing automated tools for the open-source ecosystem, has never been studied. We believe the research community can invest in mining the marketplace repository to gain more insight into different interactions and requirements within the open-source software community.

**Second,** we presented the results of a systematic mapping study of the literature in automation within the open-source community. We mapped the studies for a side-by-side comparison of academic and practical production tools in open source. There are multiple studies in the context of open source [42][19][9]. However, ours particularly provides a mapping to the perspective as seen officially by GitHub and the open-source developers.

**Third,** we performed a comprehensive comparison between our findings from mining the marketplace and performing the mapping study. This comparison of state-of-the-art and state-of-the-practice demonstrates and quantifies the difference between the research perspective and practitioners' perspective for automation in the open-source ecosystem.

Our findings provide insight into the current landscape of automation tools for open-source software, particularly within the context of GitHub Marketplace which is crucial for both researchers and practitioners in the field of software engineering and automation. Furthermore, the quantitative and qualitative analysis of GitHub Marketplace can guide future research efforts by highlighting areas of interest or potential areas for improvement in the development and adoption of automation tools.

### 5.1. The Relation Between Marketplace Features and Characteristics

In sections 4.1 and 4.2, we described various attributes of the GitHub Marketplace and classified them as Characteristics or Features and Policies following the framework of Jansen and Bloemendal [35]. Additionally, we divided the attributes of the marketplace into Developer, Platform, and User attributes (Figure 2) following Martin et al. [47].

To further investigate the potential impact of the GitHub Marketplace, we performed a heuristic search to mine confounding factors between Characteristics, Features, and Policies of the GitHub Marketplace [57, 4, 38, 90] and calculated the correlation between the Features and Characteristics of this marketplace. These correlation matrices are depicted in Figure 12. We did not identify a strong correlation between any two attributes in the marketplace. The number of contributors to Actions shows a weak correlation with factors like the number of stars, verified creator status, and Action releases. This implies that larger teams tend to be verified, release more versions of Actions, and produce more popular Actions. This prompts an inquiry into whether team size or the involvement of specific individuals drives these outcomes.

GitHub verification is one of the prominent policies in the marketplace (See Section 4.2). We observed weak correlations between GitHub Verification and key attributes such as the presence of a free trial, documentation, and verified domains in Apps. These align with factors considered in GitHub's verification process, many of which are also addressed in our study as Features and Policies. We also did not observe any strong correlation between GitHub Verification and the popularity of the tools. We observed that 66.8% (250) of the 374 Apps are not verified by the marketplace (Feature and Policies). These Apps, on average, are installed 436.7 times, while the ones verified by the marketplace have an average install of 21,831.36 times. A similar phenomenon holds for the verified domain (another example of Features and Policies), where Apps with at least one verified domain have significantly higher average installs than those without verified domains (248 Apps with an average installs of 2,336.41). We also observed the same occurrence for the Actions on the GitHub Marketplace, where the average number of stars for Actions with verified creators (2.67% of Actions) is higher than those without a verified creator (115 versus 17.35, respectively). Overall, having GitHub verification and verified domains (both examples of the Features and Policies of the platform) positively impacts the popularity of the tool (an example of platform Characteristics).

*5.2. Marketplaces Facilitate Mobilization and Innovation*

We believe that by directing our efforts toward leveraging the marketplace, we can effectively facilitate knowledge mobilization. The analysis and mining of the GitHub Marketplace (such as the one done in our study) can *provide researchers with access to a diverse set of perceptions*. A deep look into the developers' ways of automating their tasks on repositories and observing their actions can further reduce the bias in questioning and increase our leverage in narrowing the gap.

The community has known and acknowledged the gap between academia and industry. Lo et al. [44] gathered 571 Papers from ICSE and ESEC/FSE venues and asked practitioners to annotate their relevancy. The results showed that for 71% of all ratings, the publications were considered essential or worthwhile while emphasizing the room to improve the relevance and actionability of the research. This study was later replicated by Franch et al. [18] and showed a similar trajectory. On the other hand, Devanbu et al. [14] invested and demonstrated that developers often hold strong a priori opinions about several technical aspects unsupported by evidence. Such beliefs formed by subjective understanding and personal experience are error-prone. In a follow-up article [15], they echoed the concern of many researchers, "Our life's work, embodied in research papers, counted so little toward forming the opinions of professional practitioners at one of the world's leading software companies?". Further, a more recent study by Shrikanth et al. [88] pointed to the disconnection between the beliefs of practitioners and what we achieve with empirical research. The authors encourage practitioners and researchers to reassess their assumptions regularly, ensuring alignment with evolving technologies. This approach not only strengthens the relevance of research to practical applications but also enhances industrial competitiveness by leveraging cutting-edge technologies.

The community emphasizes performing relevant research for practitioners. The lack of access to data from industrial players has made the open-source community and projects to be the subject of many studies in the field, hoping that working on these open-source projects has at least gotten us closer to real-world issues. Our analysis of **RQ3** does indeed show the gap between the re-searchers' effort and the practitioners' work in the open-source community. Open source is a well-known innovation model in the software ecosystem [52, 69, 51]. In a comparative study between the open-source and commercial models, Smith [89] argues that marketplaces are critical to fostering software innovation and growth. He concludes that users benefit from increased choices and competition in such marketplaces, establishing a stronger relationship between academic work and the tools in the marketplace. The GitHub Marketplace provides a unique opportunity for the distribution and dissemination of production tools. The platform's openness, along with the flexible policies we discussed and the potential to access a wider range of practitioners and real-world scenarios, provides a unique opportunity for accelerating technological innovation through academic research.

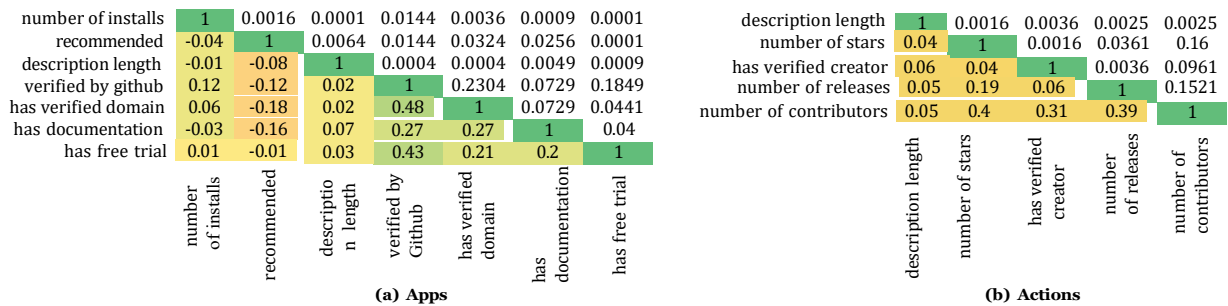

Figure 12: Correlation matrix between different attributes for (a) Apps and (b) Actions. The upper triangle shows the effect size. As effect size, we use the coefficient of determination (goodness of fit).

## 6. Threat to Validity

This study attempts to obtain a picture of the current state of research and practice of open-source software production tools. We have designed measures to elaborate on the state of the art and practice and the mobilization gap.
These measures, however, have some threats to validity that we discuss below. We refer to the validity types introduced by Wohlin et al. [93] to structure these limitations.
As for the *conclusion validity*, the threat of drawing wrong conclusions for **RQ1** and **RQ2** is low as the nature of questions is descriptive. We used conventional statistical testing and visualizations to summarize our observations. We used categories as the main unit for our side-by-side comparison. The manual categorization is fuzzy and one contribution (either the papers or Apps and Actions) can be categorized in multiple areas. We used a strict protocol with multiple annotators to reduce this risk. We used ratings and the number of stars to analyze the popularity of Apps and Actions in the marketplace (**RQ1**). The inaccuracy of these indicators has been discussed by Ruiz et al. [50] for mobile apps. Also, Borges et al. [6] warned about relying on the number of stars as a metric for repository quality.

When it comes to our diagnostic evaluation in **RQ3**, there are several potential threats. We used the number of papers and authors, as well as the number of Apps, Actions, and their developers as a proxy for the extent of interest from the community. However, these measures might not be an accurate proxy. To mitigate the risk, we used the average number of citations among all the papers and the average number of installs and stars for Apps and Actions within each category.

As per the *construct validity*, the Exclusion Criteria applied to the systematic mapping may have caused us to ignore some relevant papers in the field. Yet, we believe the search was extensive enough to cover important aspects for our comparison. We also used the categories on the GitHub Marketplace to categorize the 515 Papers from our dataset. However, the academic and industry terminologies might not be consistent. To mitigate this risk, before mapping the studies into the categories in **RQ3**, we familiarized ourselves with the official category definitions in the marketplace. We looked into samples of Apps and Actions within a category and discussed them within the group to have a similar understanding of the category definitions. Further, we used NLP techniques on the App and Action descriptions to better understand categories for our qualitative comparison in **RQ3**. While this was not good enough to allow us to automate the categorization process, it gave us a better idea of what the categories meant in the context of the GitHub Marketplace. For the final categorization, two paper authors went through all the products manually and finalized the categorization. This process could be biased, but the level of disagreement was within an acceptable range.

When it comes to the *internal validity*, in **RQ3**, we categorized the papers manually. Although the process was independent and followed by the discussion of disagreements through a mediator, bias, and misunderstandings could still threaten internal validity. We also used a combination of an automatic and a manual process to identify individual researchers. We used a combination of names, surnames, and affiliations to identify the individual authors in the field. We followed the process by manually inspecting scholar profiles and LinkedIn accounts. Yet, there is a small chance that some individuals were not identified properly, and a few redundancies exist.

As per the *external validity*, for the academic studies (**RQ3**), the number of citations over time was available. However, the marketplace does not show Apps and Actions that were published or the change of rating over time. As a result, we could not perform a similar time-based analysis or any analysis about the evolution of categories over time.

# 7. Conclusion and Future studies

GitHub Marketplace is a software repository that provides tools for the automated production of code hosted on GitHub. The marketplace provides a plethora of information about the tools available on this platform. We discussed the Characteristics, Features, and Policies of the GitHub Marketplace by mining and analyzing GitHub Marketplace's Provider, User, and Platform attributes both quantitatively and qualitatively.

We also presented systematic mapping of research studies on automation tools for open-source software. In our review of 515 papers, we categorized them according to the types of research and analytical tools used, as well as the tasks they addressed. We quantified and highlighted the gaps, helping researchers and practitioners identify opportunities and synergies to work toward reducing the gap.

We also found that although some of the automation topics in literature are widely used in practice, they have yet to align with the state of practice. Further, we observed that in the marketplace, the percentage of automation tools relevant to 'Code quality' is less than one-fifth of the papers published in this field. We found that there is significant research work on topics such as 'Testing', 'Code quality', and 'Continuous integration' while practitioners are still not making full use of them. A time series analysis of the number of published papers and created tools on a certain category (or topic) could shed some light on the nature of this misalignment between state-of-art and practice.

The GitHub Marketplace, a relatively unexplored software repository in software engineering research similar to mobile app stores, serves as a dual-sided platform for software developers and users, offering a blend of qualitative and quantitative data. This presents a significant opportunity for researchers to delve into new dimensions of software development practices and tools, potentially bridging gaps in empirical understanding. Recent studies underscore a persistent gap between academic research and industry practice, with efforts like GHTorrent facilitating access to open-source data but often lacking direct relevance assessment to practitioners' needs. Leveraging the GitHub Marketplace for repository mining provides insights into developers' automation practices and behaviors, promoting the integration of research tools into real-world applications. By encouraging academia to publish tools in the marketplace, we can enhance research impact, reproducibility, and collaboration with industry, thus advancing the practical application of software engineering research.

# Acknowledgments

This research is supported by NSERC Discovery Grant RGPIN-2019-05697.

# Appendix 1: List of the Mapped Studies

**Appendix 2: Categorized Papers**

The Table .4 contains the mapped studies into each category of the GitHub marketplace. For answering **RQ3**, two authors manually mapped the studies into the schemas introduced in the paper.

Table .4: Papers mapped to GitHub Marketplace categories

| Category | ID |
| --- | --- |
| API management/Checking | P13, P22, P52, P126, P172, P245, P322, P340, P372, P383, P454, P455, P498 |
| Testing | P11, P15, P21, P49, P59, P64, P67, P71, P76, P78, P79, P82, P94, P101, P114, P116, P117, P118, P123, P124, P146, P150, P154, P156, P159, P162, P163, P165, P171, P174, P178, P179, P189, P195, P203, P207, P208, P212, P234, P244, P246, P247, P248, P250, P257, P260, P261, P263, P266, P267, P268, P272, P286, P305, P313, P323, P330, P337, P339, P342, P343, P348, P352, P353, P354, P357, P361, P362, P363, P364, P380, P384, P398, P417, P425, P426, P428, P432, P437, P438, P441, P442, P444, P450, P458, P459, P462, P464, P466, P470, P473, P474, P476, P480, P483, P487, P489, P492, P493, P496, P497, P502, P510, P515 |
| Utilities | P8, P10, P28, P30, P34, P41, P51, P52, P61, P87, P90, P95, P103, P106, P116, P125, P126, P136, P137, P140, P141, P145, P149, P150, P152, P154, P158, P161, P175, P180, P181, P193, P216, P221, P243, P251, P254, P284, P322, P349, P394, P397, P403, P405, P418, P448, P463, P468, P494 |
| Reporting | P16, P35, P36, P40, P53, P55, P57, P59, P69, P86, P87, P92, P96, P160, P166, P168, P202, P206, P219, P224, P225, P271, P309, P316, P330, P343, P350, P351, P352, P356, P413, P418, P468, P477, P484, P493, P503 |
| Continuous integration | P29, P71, P85, P180, P210, P348, P350, P353, P368, P369, P374, P376, P385, P386, P388, P480 |
| Publishing | P6, P19, P54, P85, P107, P172, P180, P201, P223, P227, P273, P275, P276, P287, P288, P296, P335, P336 |
| Support | P8, P12, P18, P21, P34, P38, P45, P60, P65, P68, P70, P72, P74, P85, P88, P92, P94, P103, P105, P107, P112, P121, P143, P154, P155, P156, P161, P183, P193, P201, P202, P214, P225, P227, P233, P249, P251, P256, P259, P270, P274, P278, P284, P288, P294, P310, P326, P359, P388, P406, P408, P424, P428, P431, P434, P439, P441, P454, P460, P461, P468, P471, P472, P475, P479, P488, P490, P491, P504, P506, P512 |
| Project management | P5, P6, P9, P12, P16, P19, P29, P37, P47, P57, P75, P102, P107, P113, P121, P134, P138, P144, P168, P173, P223, P229, P233, P236, P240, P241, P243, P249, P253, P255, P262, P270, P273, P274, P275, P276, P281, P283, P296, P304, P312, P316, P325, P326, P332, P335, P336, P354, P362, P366, P373, P374, P404, P405, P418, P457, P460 |
| Code review | P1, P2, P17, P22, P28, P41, P48, P56, P58, P66, P68, P70, P80, P86, P91, P93, P98, P103, P108, P111, P119, P126, P130, P141, P142, P151, P152, P173, P191, P215, P218, P232, P235, P273, P282, P297, P300, P345, P347, P349, P352, P356, P357, P358, P359, P360, P364, P365, P367, P368, P372, P374, P375, P377, P378, P379, P381, P387, P396, P416, P420, P421, P436, P451, P452, P453, P454, P471, P487, P491 |
| Deployment | P9, P67, P149, P175, P275, P276, P294, P308 |
| Chat | P27, P31, P62, P206, P221, P382, P389 |
| Community | P5, P19, P20, P27, P29, P31, P37, P40, P48, P60, P61, P62, P81, P100, P104, P105, P120, P227, P259, P288, P346, P360, P366, P370, P377, P381, P408, P427 |
| Dependency management | P7, P66, P129, P145, P147, P157, P166, P179, P182, P187, P188, P201, P207, P211, P213, P215, P220, P228, P237, P238, P239, P388, P409, P413, P419, P463, P484, P505 |
| AI Assisted | P9, P14, P17, P28, P46, P50, P53, P63, P69, P79, P89, P99, P113, P122, P128, P132, P133, P139, P141, P143, P144, P147, P149, P152, P157, P168, P181, P198, P211, P212, P213, P217, P220, P229, P233, P234, P235, P236, P237, P238, P239, P240, P246, P250, P274, P281, P283, P290, P294, P299, P307, P318, P328, P334, P345, P346, P355, P356, P358, P363, P367, P383, P390, P394, P397, P399, P400, P401, P402, P403, P405, P415, P423, P429, P435, P456, P461, P474, P485, P495, P498, P503, P508, P513 |
| Open Source management | P11, P22, P37, P47, P60, P61, P62, P65, P71, P72, P81, P95, P104, P105, P134, P206, P209, P221, P229, P360, P367, P371, P387, P391, P404, P427, P504 |
| Security | P45, P122, P134, P158, P177, P182, P187, P263, P282, P298, P340, P347, P351, P354, P357, P359, P369, P373, P380, P410, P435, P440, P443, P446, P449, P469, P476, P477, P486, P496, P499, P501 |
| Monitoring | P68, P97, P102, P110, P118, P125, P129, P137, P138, P144, P153, P157, P160, P163, P182, P236, P249, P254, P262, P308, P378, P392, P393, P395, P424, P465, P494 |
| Code quality | P1, P2, P13, P21, P35, P49, P51, P56, P58, P63, P66, P76, P77, P80, P82, P84, P88, P89, P90, P91, P93, P96, P98, P99, P101, P106, P108, P110, P111, P113, P115, P119, P123, P125, P128, P129, P130, P131, P132, P133, P136, P137, P139, P140, P142, P143, P146, P151, P153, P156, P158, P164, P166, P167, P170, P173, P174, P175, P179, P181, P189, P190, P192, P200, P203, P209, P210, P214, P215, P216, P223, P232, P238, P242, P250, P252, P256, P257, P262, P263, P265, P271, P272, P277, P285, P289, P292, P295, P296, P297, P300, P305, P312, P314, P327, P331, P344, P345, P347, P351, P353, P355, P361, P364, P365, P372, P375, P376, P379, P383, P385, P387, P411, P414, P415, P416, P422, P430, P433, P434, P436, P437, P438, P441, P445, P447, P467, P470, P471, P477, P478, P481, P482, P485, P506, P507, P508, P511, P514, P515 |
| Localization | P13, P131, P132, P133, P136, P147, P188, P207, P211, P213, P220, P237, P239, P371, P412 |
| Desktop tools | P4, P87, P95, P155, P225 |
| Mobile | P40, P64, P67, P96, P114, P169, P244, P439, P443 |
| IDEs | P34, P41, P191, P349, P361, P431, P448, P465, P472, P509 |
| Mobile CI | P64 |
| Code search | P1, P7, P11, P17, P20, P35, P52, P56, P58, P77, P80, P82, P88, P91, P93, P99, P101, P106, P111, P119, P123, P139, P140, P142, P150, P151, P171, P183, P195, P198, P224, P248, P252, P256, P257, P285, P355, P358, P363, P407, P461, P464, P472, P487, P491, P500 |
| Code scanning ready | P2, P77, P89, P128, P131, P188, P195, P202, P224, P251, P272, P282, P365, P369, P376, P378, P379, P380, P463, P464, P508 |
| Learning | P5, P36, P108, P112, P160, P191, P259, P278, P309, P343, P366, P368, P370 |
| Time tracking | P14, P47, P138, P235 |

47